\documentclass[twocolumn,aps,prl,superscriptaddress,showpacs]{revtex4-1}
\usepackage{amssymb}
\usepackage{mathrsfs}
\usepackage{lipsum}
\usepackage{graphicx}
\usepackage[caption=false]{subfig}
\usepackage{mathtools}
\usepackage{epstopdf}
\usepackage{xcolor}
\definecolor{ora}{RGB}{180,0,160}
\usepackage[colorlinks, linkcolor=red, anchorcolor=green, citecolor=blue, urlcolor=ora]{hyperref}
\DeclareGraphicsExtensions{.pdf,.jpg,.png,.eps}
\usepackage[toc,page,title,titletoc,header]{appendix}

\begin{document}
	
\title{Control of Structured Light Enables Nearly Perfect Noise-filtering}

\author{Jian-Dong Zhang}
\affiliation{School of Physics, Harbin Institute of Technology, Harbin 150001, China}
\author{Zi-Jing Zhang}
\email[]{Authors to whom any correspondence should be addressed. \\  zhangzijing@hit.edu.cn}
\affiliation{School of Physics, Harbin Institute of Technology, Harbin 150001, China}
\author{Long-Zhu Cen}
\affiliation{School of Physics, Harbin Institute of Technology, Harbin 150001, China}
\author{Bin Luo}
\email[]{luobin@bupt.edu.cn}
\affiliation{State Key Laboratory of Information Photonics and Optical Communications, Beijing University of Posts and Telecommunications, Beijing 100876, China}
\author{Chenglong You}
\affiliation{Department of Physics and Astronomy, Louisiana State University, Baton Rouge, Louisiana 70803, USA}
\author{Omar~S.~Maga\~na-Loaiza}
\affiliation{Department of Physics and Astronomy, Louisiana State University, Baton Rouge, Louisiana 70803, USA}
\author{Yi-Fei Sun}
\affiliation{School of Physics, Harbin Institute of Technology, Harbin 150001, China}
\author{Lu Xu}
\affiliation{School of informatics, Zhejiang Sci-Tech University, Hangzhou 310018, China}
\author{Long Wu}
\affiliation{School of informatics, Zhejiang Sci-Tech University, Hangzhou 310018, China}
\author{Yuan Zhao}
\email[]{zhaoyuan@hit.edu.cn}
\affiliation{School of Physics, Harbin Institute of Technology, Harbin 150001, China}

\date{\today}
	
\begin{abstract}
	The performance of laser-based active sensing has been severely limited by two types of noise: electrical noise, stemming from elements; optical noise, laser jamming from an eavesdropper and background from environment.
	Conventional methods to filter optical noise take advantage of the differences between signal and noise in time, wavelength, and polarization.
	However, they may be limited when the noise and signal share the same information on these degrees of freedoms (DoFs).
	In order to overcome this drawback, we experimentally demonstrate a groundbreaking noise-filtering method by controlling orbital angular momentum (OAM) to distinguish signal from noise.
	We provide a proof-of-principle experiment and discuss the dependence of azimuthal index of OAM and detection aperture on signal-to-noise ratio (SNR).
	Our results suggest that using OAM against noise is an efficient method, offering a new route to optical sensing immersed in high-level noise.
\end{abstract}

\maketitle

{\emph{Introduction.---}}
Since the invention of lasers, laser-based active sensing has been playing an important role in a large range of fields, such as imaging \cite{Gariepy2015Single,Shin2016Photon}, communication \cite{Ma2012Quantum,Krenn13648}, detection \cite{PhysRevLett.108.183901,Lavery537}, and metrology \cite{RevModPhys.84.777}.
But various noise in the system puts a spanner when it comes to practical applications. 
The noise degrades the performance of sensing, and may even make sensing unable to implement when the noise level is higher than signal.
Therefore, protecting signal from noise is a long-lasting subject.
In general, noise in a sensing system can be divided into two categories: electrical noise and optical noise.
The former is induced by electrical elements, including shot noise, generation-recombination noise, thermal noise, temperature noise, and flicker noise.
This kind of noise can be suppressed by optimizing circuit design and selecting low-noise elements.
There are two kinds of optical noise, laser jamming noise from an eavesdropper, and background noise from the environment.
The task of reducing the effects arising from aforementioned optical noise has gained lots of attention recently. 
Here, jamming and background are the types of noise about which we are concerned.

So far, conventional methods for filtering optical noise are mainly based upon the differences between signal and noise in some DoFs.
The original quantum illumination protocol proposed by Lloyd used a pair of frequency-entangled photons to suppress high background noise \cite{Lloyd1463}.
Related to this, some quantum illumination protocols were demonstrated, which can reduce noise through the use of temporal or spatial correlations \cite{PhysRevA.99.023828,PhysRevLett.110.153603,gregory2019imaging}.
Malik \emph{et al.} reported a quantum-secured imaging protocol that can discriminate intercept-resend jamming attack using photon's polarization DoF \cite{doi:10.1063/1.4770298}.
Morris \emph{et al.} developed a low-light imaging protocol in which background counts can be filtered with time-gated technique \cite{Morris2015Imaging}.
More recently, a long-range three-dimensional imaging was achieved by Li \emph{et al.}, background noise can be filtered with polarized and temporal DoFs \cite{li2019single}.  
Cohen \emph{et al.} presented a threshold quantum LiDAR which SNR can be improved by exploiting the difference between photon number distribution of signal and that of noise \cite{cohen2019thresholded}.
However, due to indistinguishability, these methods have no way to deal with noise that is the same as signal regarding time, frequency, polarization, and photon number distribution. 
As a consequence, an efficient noise-filtering method is needed to bridge this gap.

The limited performance of these methods originates from the fact that the noise generally pervades entire interval of the selected DoF.
A reliable method is to exploit a unique DoF of which noise is located in a specific interval. 
Due to the absence of innate high-dimensional OAM strcuture in noise, in this Letter, we propose a noise-filtering method by adding OAM into signal. 
With the OAM introduced, the spatial distribution of signal is different from that of noise. 
Further, they can be spatially separated through the use of phase modulation. 
Our method is capable of filtering noise that has DoFs---time, frequency, polarization, and photon number distribution---in common with signal; moreover, it can assist conventional methods to further improve SNR.
This provides a new route to areas that need to extract signal from noise, and guarantees the implementation of sensing in the presence of high-level noise.

\begin{figure}[htbp]
	\centering
	\includegraphics[width=.48\textwidth]{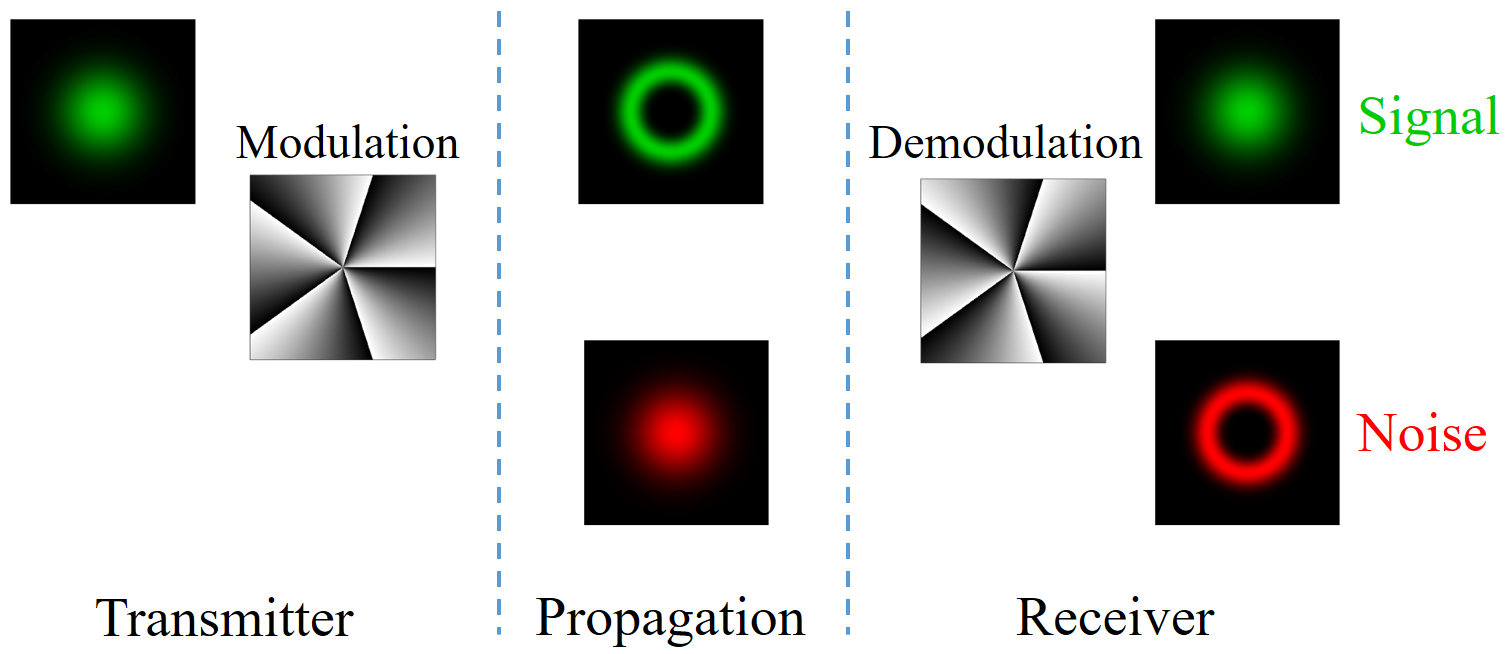}
	\caption{The principle of noise-filtering based on spatial distribution difference between signal and noise. Top and bottom are intensity profiles of noise and signal at each stage, respectively. Two reverse spiral phase are used in modulation and demodulation. }
	\label{f0}
\end{figure}

{\emph{Fundamental principle.---}}
In what follows, we illustrate the fundamental principle of our method for filtering both spatial coherence and incoherence noise, jamming and background. 
As shown in Fig. \ref{f0}, signal is transmitted after spiral phase modulation, which differs from usual sensing schemes.
Based on high-probabilitic approximation \cite{DENNIS2009293,note1}, the signal during the process of propagation can be written as a Laguerre-Gaussian (LG) mode \cite{PhysRevA.45.8185}. 
The phase term, $\exp(i\ell\theta)$ with $\ell$ being azimuthal index of OAM, creates a dark core in the center of intensity profile.
At the receiver, it is offset by demodulation generated by a reverse spiral phase, $\exp(-i\ell\theta)$; consequently, the signal convergences to a nearly Gaussian mode as the propagation distance increases.
On the other hand, the jamming noise and background noise propagate without spiral phase modulation.
Accordingly, their intensity profiles are found to be Gaussian at the receiver.
On completion of the spiral phase demodulation, intensity profiles of the jamming noise and background noise evolve into vortices with a central dark core.
Hence, the signal sits at the center of the vortices generated by noise.
This suggests that SNR can be improved by only detecting the center, as intensity profiles of signal and noise in this region are bright and dark cores, respectively.

\begin{figure}[htbp]
	\centering
	\includegraphics[width=.48\textwidth]{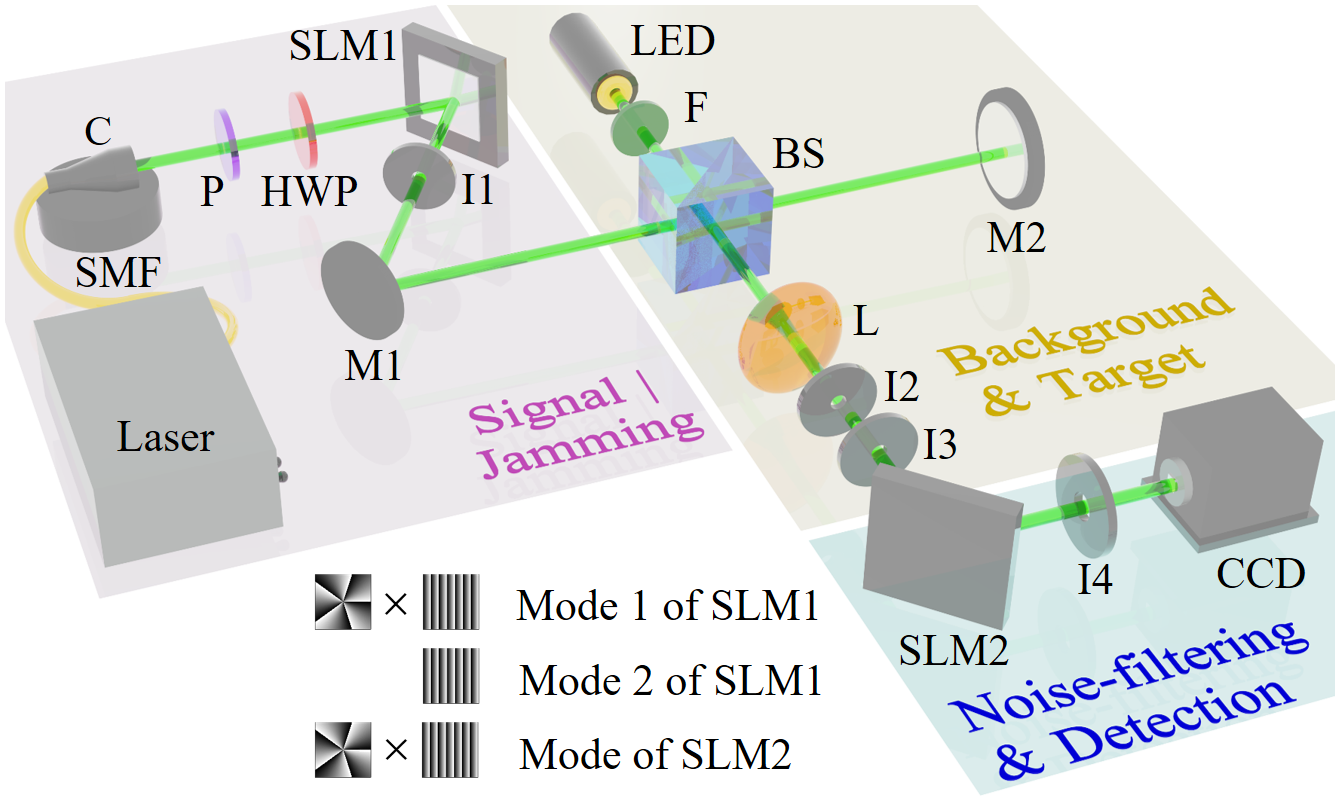}
	\caption{A schematic layout of the experiment. The abbreviations are defined as follows: SMF, single-mode fiber; C, collimator; P, polarizer; HWP, half-wave plate; SLM, spatial light modulator (HOLOEYE, LETO-VIS-009); I, iris; M, mirror; BS, beam splitter; LED, light emitting diode; F, filter; L, lens;  CCD, charge coupled device (THORLABS, 8051M-USB-TE).}
	\label{f1}
\end{figure}

Based on this principle, we perform a proof-of-principle experiment to demonstrate the advantages for improving SNR.
Consider the experimental setup of our noise-filtering method, as depicted in Fig. \ref{f1}. 
A beam from a pulsed laser is coupled into a single-mode fiber to clean its spatial mode. 
The polarization of the signal is prepared by a polarizer and a half-wave plate. 
Subsequently, the beam is incident on the first phase-only spatial light modulator (SLM).
With mode 1 chosen, the SLM shows a $\ell$-fold spiral phase mask along with a blazed grating; as a result, the beam carries OAM and can be seen as signal.
Regarding mode 2, there is only a blazed grating added to the SLM; accordingly; the output contains no OAM and acts as jamming noise.
An iris at the Fourier plane retains the first-order diffraction of the SLM, and a mirror changes propagation direction of the beam. 
Then a beam splitter (BS) takes the role of a versatile transceiver.
The beam upon leaving the BS illuminates a specular target.
Meanwhile, the background noise generated from an LED source is injected into the BS.
A bandpass filter in front of the LED is used to get rid of the background whose wavelength deviates from signal's wavelength.    
At the part of noise-filtering and detection, a lens is placed to control the sizes of signal and noise.
Two irises with the same radius are used to enhance the spatial coherence of background \cite{PhysRevLett.92.143905} and the purity of ultimate OAM spectrum \cite{PhysRevA.90.063801}.
As the key of noise-filtering, SLM2 displays a reverse $\ell$-fold spiral phase mask along with a blazed grating.
Finally, by picking a proper diameter of detection aperture (${\rm I}_4$), one can realize nearly perfect noise-filtering.

\begin{figure}[htbp]
	\centering
	\includegraphics[width=.45\textwidth]{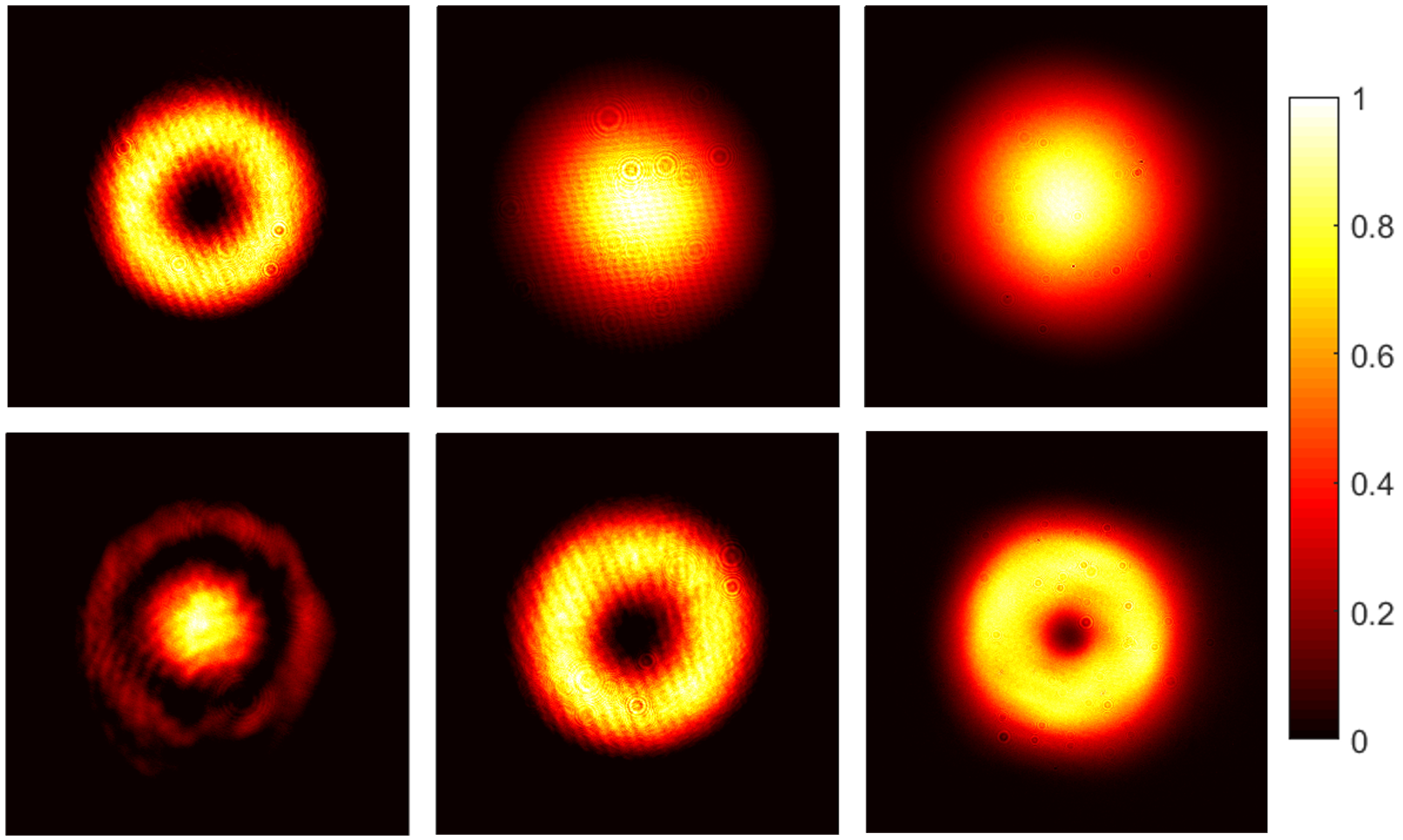}
	\caption{Top and bottom are intensity profiles, without actual scaling, before and after SLM2, respectively. From left to right: signal, jamming, and background. The color bar on the right indicates the corresponding value of the intensity.}
	\label{f2}
\end{figure}

In Fig. \ref{f2}, we show the intensity profiles of signals and noise, taken by a CCD camera.
Before SLM2, the intensity profile of signal is donut-like while those of jamming and background are Gaussian. 
Upon the illumination on SLM2, the signal evolves into Gaussian mode; meanwhile, both background and jamming produce dark cores in their intensity profiles.

It should be noted that the intensity profile of background manifests a narrower core compared to that of jamming for a fixed azimuthal index of OAM.
This originates from the difference of spatial coherence between two kinds of optical noise.
As a spatial coherent source, the mathematical description of jamming is well-studied. 
It can be viewed as an approximate LG mode after SLM2.
However, the background is a spatial incoherent source; in turn, it is by no means easy to provide a mathematically tractable expression.
An common way is to divide this source into many Huygens elementary sources \cite{PhysRevLett.92.143905,PhysRevLett.99.163901,PhysRevA.90.063801}; furthermore, each of them illuminates SLM2 with a spherical wave.
Namely, an incoherent source can be regarded as an incoherent superposition of coherent ones, which carry a random phase.
Therefore, the intensity profile behind SLM2 can be formally expressed as
\begin{eqnarray}
{I_{\rm B}} = \sum\limits_{ m} {{{\left| {{H_z}\left[ {\frac{{\exp \left( {ik{d_{ m}}} \right)}}{{{d_{ m}}}}\exp ( - i\ell \theta )} \right]} \right|}^2}},
\end{eqnarray}
where subscript ${\rm m}$ is the ${\rm m}$-th coherent source at the plane of the iris; $d_{m} = {[ {{{\left( {r_{2} - {r_3}} \right)}^2} + {{\left( {\theta_{2} - {\theta _3}} \right)}^2} + {f^2}} ]^{ {1 / 2}}}$ is Euclidean distance between the source point ($r_3$, $\theta_3$) and a point ($r_{2}$, $\theta_{2}$) on SLM2; $f$ denotes the distance  between the source plane and SLM2; ${{H_z}\left[  \cdot  \right]}$ stands for free-space Kirchhoff diffraction integral with a distance $z$.

Such incoherent superposition makes LG modes arising from independent coherent sources overlap with each other; as a consequence, the background gives a narrower dark core in contrast to the jamming (see Appendix for details).

{\emph{Results and discussions.---}}
On the basis of our method, now we demonstrate the advantages in SNR.
In our experiment, three intensity profiles are independently measured by a CCD camera.
In addition, we measure the divergence angles of jamming and background.
Based on their ratio, the intensity profile of background is scaled, since jamming and signal pass through the same path.
The corresponding total intensities are recorded as ${I}_{\rm S}$, ${I}_{\rm J}$, and ${I}_{\rm B}$.
In terms of the definition, the initial SNRs turn out to be
\begin{align}
{\rm SNR}_1 &= \frac{{{I_{\rm S}}}}{{{I_{\rm J}}}},\\
{\rm SNR}_2 &= \frac{{{I_{\rm S}}}}{{{I_{\rm B}}}}.
\end{align}
After noise-filtering devices, spiral phase demodulation and a detection aperture with diameter $D$, one can determine the remainder total intensity
\begin{eqnarray}
{I_{\rm{rem}}}=\int_0^{{D \mathord{\left/
			{\vphantom {D 2}} \right.
			\kern-\nulldelimiterspace} 2}} {\int_0^{2\pi } {{I_\kappa }\left( {r,\theta } \right)} } rdrd\theta  ,
\end{eqnarray}
where $\kappa$ takes S, J, and B.
Let us denote this double integral as ${\eta _\kappa }{I_\kappa }$ with a dimensionless factor $\eta$; in turn, these remainder total intensities can be represented as $\eta_{\rm S}{I}_{\rm S}$, $\eta_{\rm J}{I}_{\rm J}$, and $\eta_{\rm B}{I}_{\rm B}$.
Hence, the SNRs after noise-filtering can be written as ${\rm SNR}'_1 = \eta_{\rm S}{{{I_{\rm S}}}}/\eta_{\rm J}{{{I_{\rm J}}}}$ and ${\rm SNR}'_2 = \eta_{\rm S}{{{I_{\rm S}}}}/\eta_{\rm B}{{{I_{\rm B}}}}$.
Further, we introduce two enhancement factors
\begin{align}
{\cal F}_1 &= \frac{{\rm SNR}'_1}{{\rm SNR}_1}=\frac{\eta_{\rm S}}{\eta_{\rm J}},\\
{\cal F}_2 &= \frac{{\rm SNR}'_2}{{\rm SNR}_2}=\frac{\eta_{\rm S}}{\eta_{\rm B}}
\end{align}
to quantify the SNR improvement.

\begin{figure}[htbp]
	\centering
	\includegraphics[width=.48\textwidth]{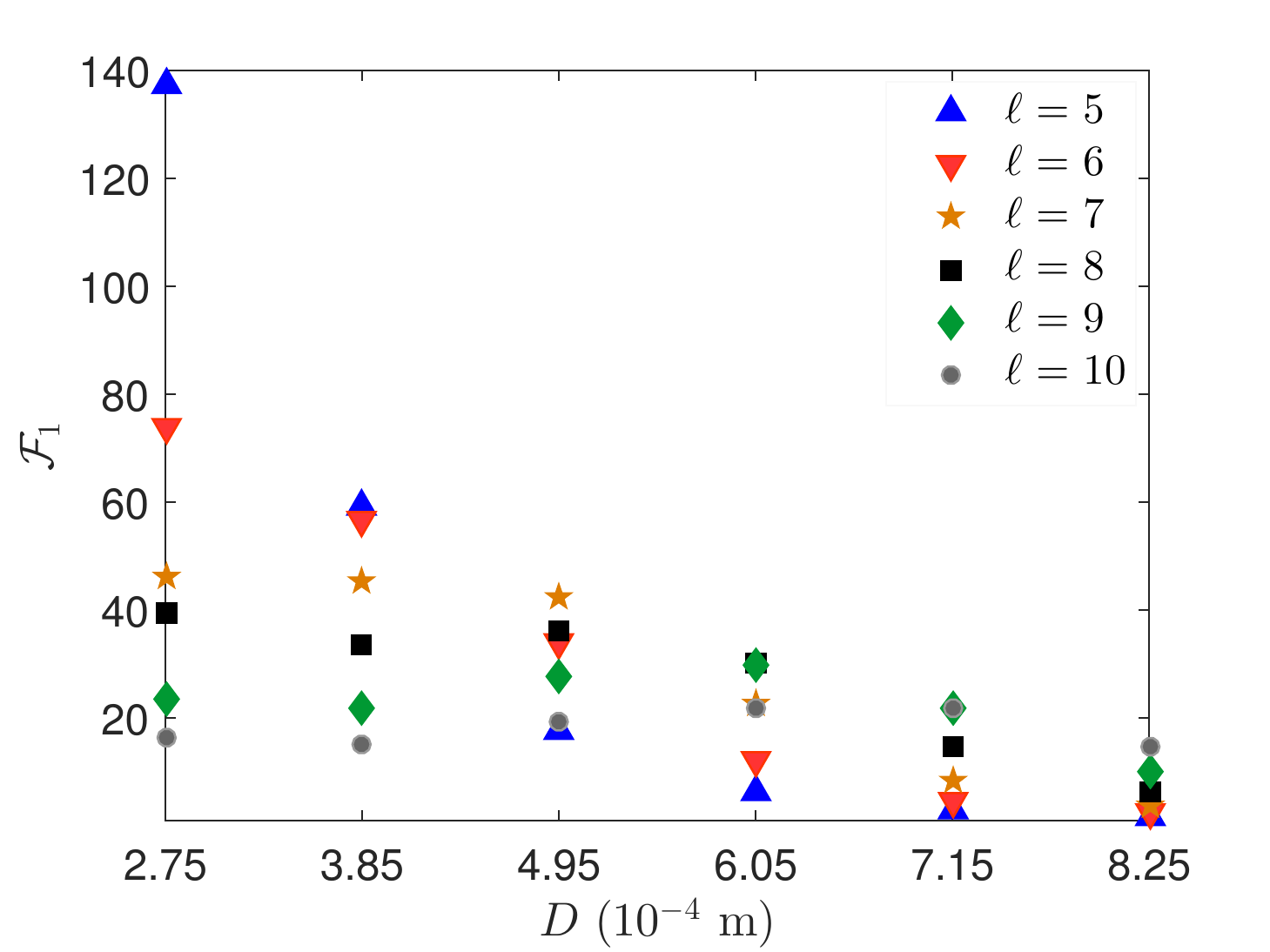}
	\caption{Enhancement factor, ${\cal F}_1$, calculated from experimental results as a function of diameter, $D$, of detection aperture. 
		The minimum value of longitudinal axis is 1, and all enhancement factors are above it.
		As the diameter decreases, the values of enhancement factors obtained by high-order azimuthal indices, $\ell=9$ and $\ell=10$, maintain around 20, while those obtained by low-order azimuthal indices, $\ell=5$ and $\ell=6$, are monotonically increasing functions.}
	\label{f3}
\end{figure}

In order to demonstrate the advantage of our method against jamming, with different azimuthal indices, $\ell$, we show the enhancement factor, ${\cal F}_1$, as a function of diameter, $D$, of detection aperture in Fig. \ref{f3}.
To every enhancement factor there corresponds a number greater than 1, which means that the advantage of our method holds true for any azimuthal indices.
Furthermore, a low-order azimuthal index comes across as an ideal candidate for the jamming, e.g., the SNR improvement with $\ell=5$ and that with $\ell=10$.
The reason behind this phenomenon is that a high-order LG mode needs a longer propagation distance and has a larger radius \cite{zhang2019angular}.
Thus, for a given diameter, we can detect more intensity with deploying a low-order azimuthal index, rather than a high-order one.

\begin{figure}[htbp]
	\centering
	\includegraphics[width=.48\textwidth]{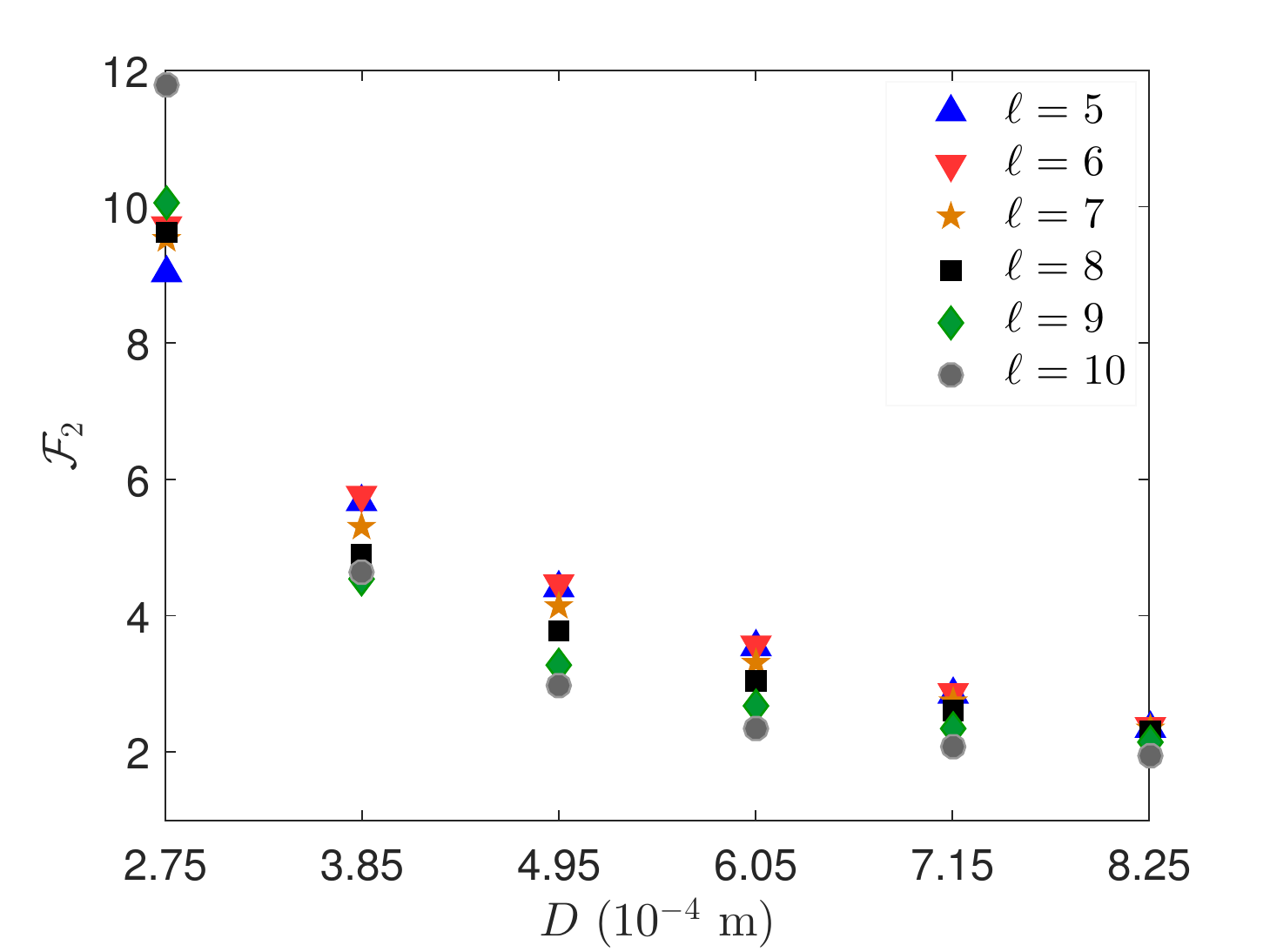}
	\caption{Enhancement factor, ${\cal F}_2$, calculated from experimental results as a function of diameter, $D$, of detection aperture.
		The minimum of longitudinal axis is 1, and all enhancement factors are above it.
		As the diameter decreases, the values of enhancement factors obtained by any azimuthal indices are monotonically increasing functions.}
	\label{f4}
\end{figure}

Regarding the advantage in filtering background, Fig. \ref{f4} provides the dependence of enhancement factor, ${\cal F}_2$, on diameter, $D$, of detection aperture.
One can find that the advantage remains as all enhancement factors are above 1.
That is, however great the size of detection aperture, our method can always give SNR improvement.
In addition, for a small diameter, the improvement is remarkable when a high-order azimuthal index is used.
This results from the fact that a incoherent source leads to incomplete destructive interference in central region; consequently, the extinction ratio with a high-order azimuthal index is superior to that with low-order ones.
Even so, all azimuthal indices are capable of improving the SNR with the decrease of diameter.

Combining Figs. \ref{f3} and \ref{f4}, it can be seen that SNR improvements reach up to one to two orders of magnitude when compared to non-OAM methods.
In contrast, the filtering ability of our method to jamming outperforms that to background because of the difference in their spatial coherence.
Indeed, we can achieve more significant improvement by further decreasing the diameter of detection aperture.
However, the signal itself faces a sharp decrease as the SNR improves.
This will produce lower detection rate; in turn, signal is at risk of being submerged in electrical noise, like dark counts.
As a consequence, in practical scenarios, we need to deal with the trade-off between SNR improvement and detection rate. 
The results reveal that the use of a lower-order azimuthal index may be a better choice, e.g., the efficiency $\eta_{\rm S}$ with $\ell=5$ is in excess of 19\% when $D \geqslant 2.75\times 10^{-4}$ m.

In addition, we would like to note three more points about our method. 
Although we use a specular target in our proof-of-principle experiment, the method we proposed can also be implemented in practical scenarios, target scattering \cite{Lavery537}, obstacle \cite{zhang2019angular}, atmospheric turbulence \cite{PhysRevLett.94.153901}, to name a few.
This is due to the fact that part of the original azimuthal index of OAM in signal stays the same after these processes.
Moreover, our method is compatible with the conventional noise-filtering ones, for these DoFs are independent.
Furthermore, a single-mode fiber may further improve SNR by reducing jamming and background, due to that almost all noise is non-Gaussian mode.

{\emph{Conclusions.---}} 
In summary, we have experimentally demonstrated a method that permits nearly prefect noise-filtering. 
Unlike previous methods, OAM is selected as a special DoF to give rise to spatial distribution difference between signal and noise.
We can effectively separate signal and noise by taking advantage of this difference and phase modulation. 
The proof-of-principle experiment proves the efficiency and correctness of our method, and SNR improves by one to two orders of magnitude beyond those of non-OAM methods. 
Our results indicate that separating signal from noise is a new method and can be used to assist conventional methods in further filtering noise owing to the independence among DoFs.
This method combined with some post-processing algorithms might develop into a ``killer application" towards most fields that need to extract signal from high-level noise.

{\emph{Acknowledgment.---}} This work was supported by National Natural Science Foundation of China (Grant No. 61701139). 

J.-D. Z. and Z.-J. Z. contributed equally to this work.

\appendix
\section{Appendix}

\begin{figure}[htbp]
	\centering
	\includegraphics[width=.48\textwidth]{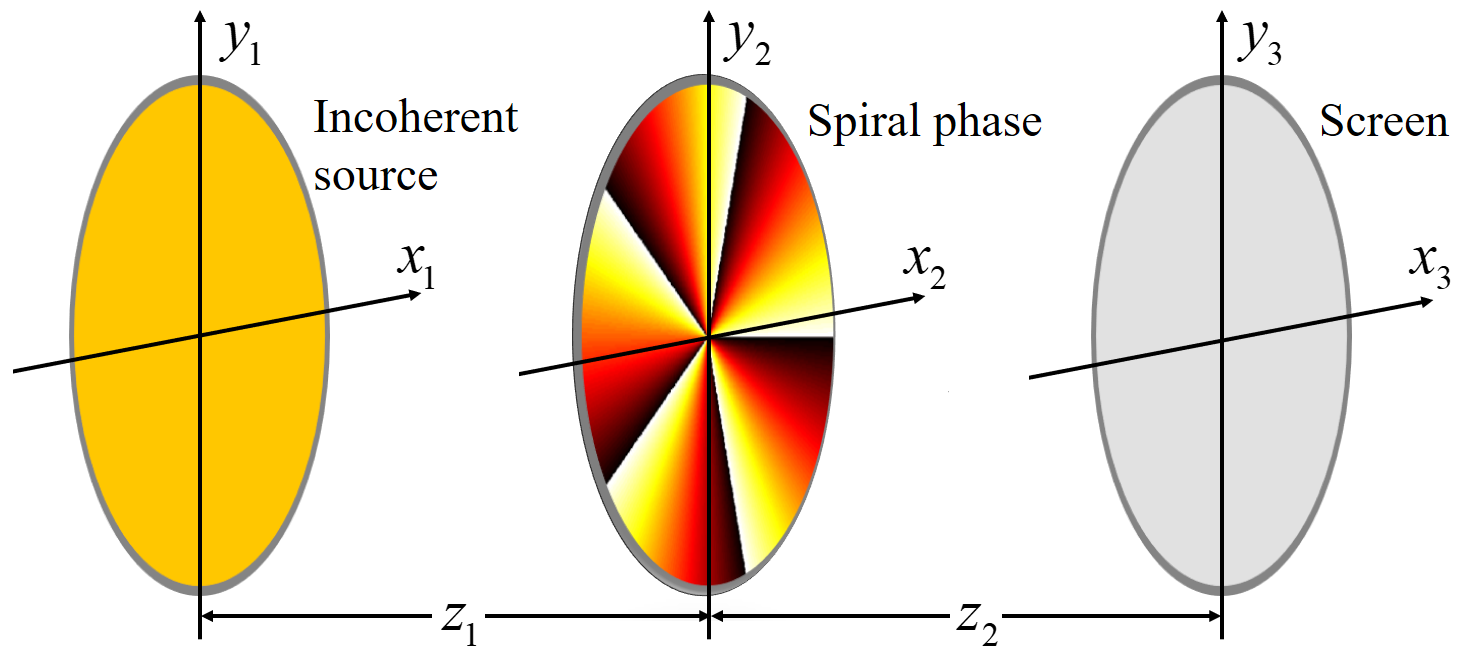}
	\caption{Illustration of characteristic analysis on dark vortex core generated from a incoherent source. The effective diameters of all planes are 1 mm.}
	\label{f5}
\end{figure}

Here we give some discussions about the properties of optical vortex generated from a incoherent source.
As illustrated in Fig. \ref{f5}, a spiral phase placed at $z_1$ plane is illuminated by a spatially incoherent source.
Then through free-space propagation, one can detect the corresponding vortex at the screen.
As mentioned in main text, a incoherent source can be regarded as a combination of many coherent ones.
Thus, we consider two scenarios: on- and off-axis coherent sources.

\begin{figure}[htbp]
	\centering
	\includegraphics[width=.48\textwidth]{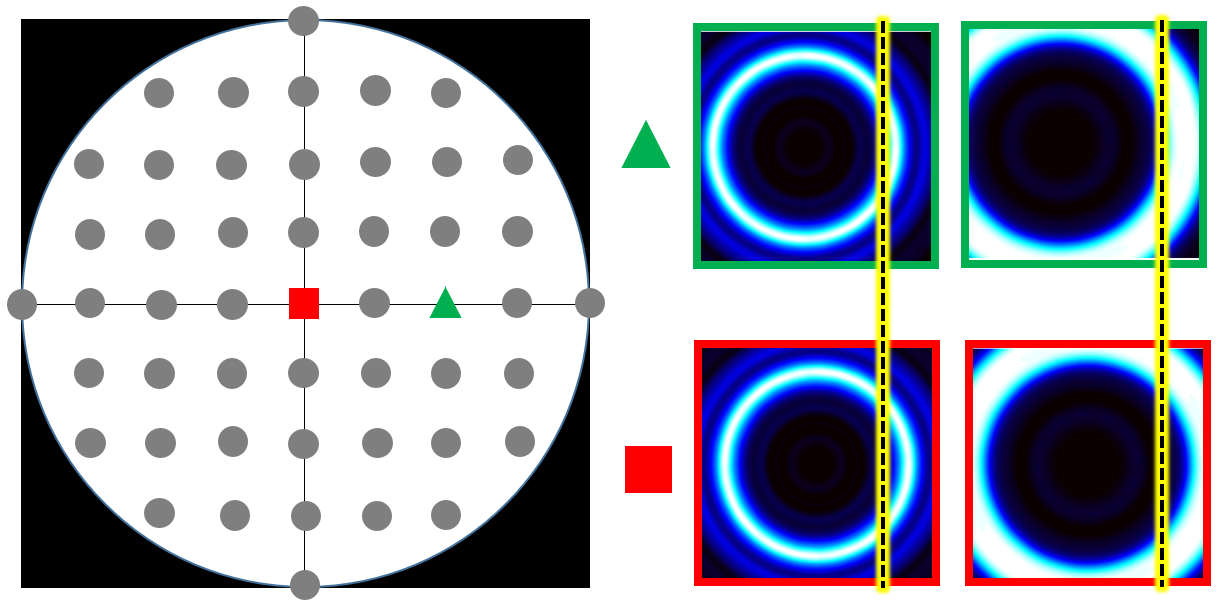}
	\caption{Intensity profiles at the screen and corresponding Huygens elementary sources. The dimensions of all intensity profiles are 1 mm $\times$ 1 mm. The displacement of   triangular point relative to the center is 0.25 mm, and $\ell=5$ and $z_1=0.5$ m are used. Two columns correspond to $z_2$ = 0.1 m and 0.2 m, respectively.}
	\label{f6}
\end{figure}

In Fig. \ref{f6}, we divide the plane of the incoherent source into a number of coherent point sources.
Square and triangular points represent on- and off-axis sources, respectively.
On the right-hand side, we provide the intensity profiles of vortices stemming from these two points. 
As can be seen from the figure, to every off-axis point in the plane of incoherent source there corresponds a off-axis vortex (dashed lines) at the screen.
Further, a perfect dark vortex core generated by the square point becomes obscure, for all off-axis vortices overlap with it.
Hence, the dark vortex core of a incoherent source is narrower than that of a coherent one.
In turn, a high-order index leads to a distinct vortex core, in that the radius of each vortex is large and the effect of overlap is weakened. 
Furthermore, the displacement of off-axis vortex diminishes with increasing the distance, $z_1$, suggesting a more visible dark core.
Equivalently, reducing the size of the incoherent source can obtain the same result.

Finally, we discuss the off-axis vortex from the view of angular spectrum theory.
In this way, the vortex can be decomposed into LG basis, as different LG modes are orthogonal.
At the plane of the screen, the field amplitude of the vortex can be written as:
\begin{eqnarray}
{E_{\rm scr}}\left( {r,\theta } \right) = \frac{1}{{\sqrt {2\pi } }}\sum\limits_\ell^{} {{a_\ell}} \left( r \right)\exp \left( {i\ell\theta } \right),
\end{eqnarray}
where the weight factor can be obtained from the discrete Fourier transformation
\begin{eqnarray}
{a_\ell}\left( r \right) = \frac{1}{{\sqrt {2\pi } }}\int_0^{2\pi } {{E_{\rm scr}}\left( {r,\theta } \right)} \exp \left( { - i\ell\theta } \right)d\theta. 
\end{eqnarray}
Further, with ${C_\ell} = \int_0^\infty  {{{\left| {{a_\ell}\left( r \right)} \right|}^2}} rdr$, the normalized probability can be determined by
\begin{eqnarray}
{P_\ell} = \frac{{{C_\ell}}}{{\sum\nolimits_\ell {{C_\ell}} }}.
\end{eqnarray}

\begin{figure}[htbp]
	\centering
	\includegraphics[width=.48\textwidth]{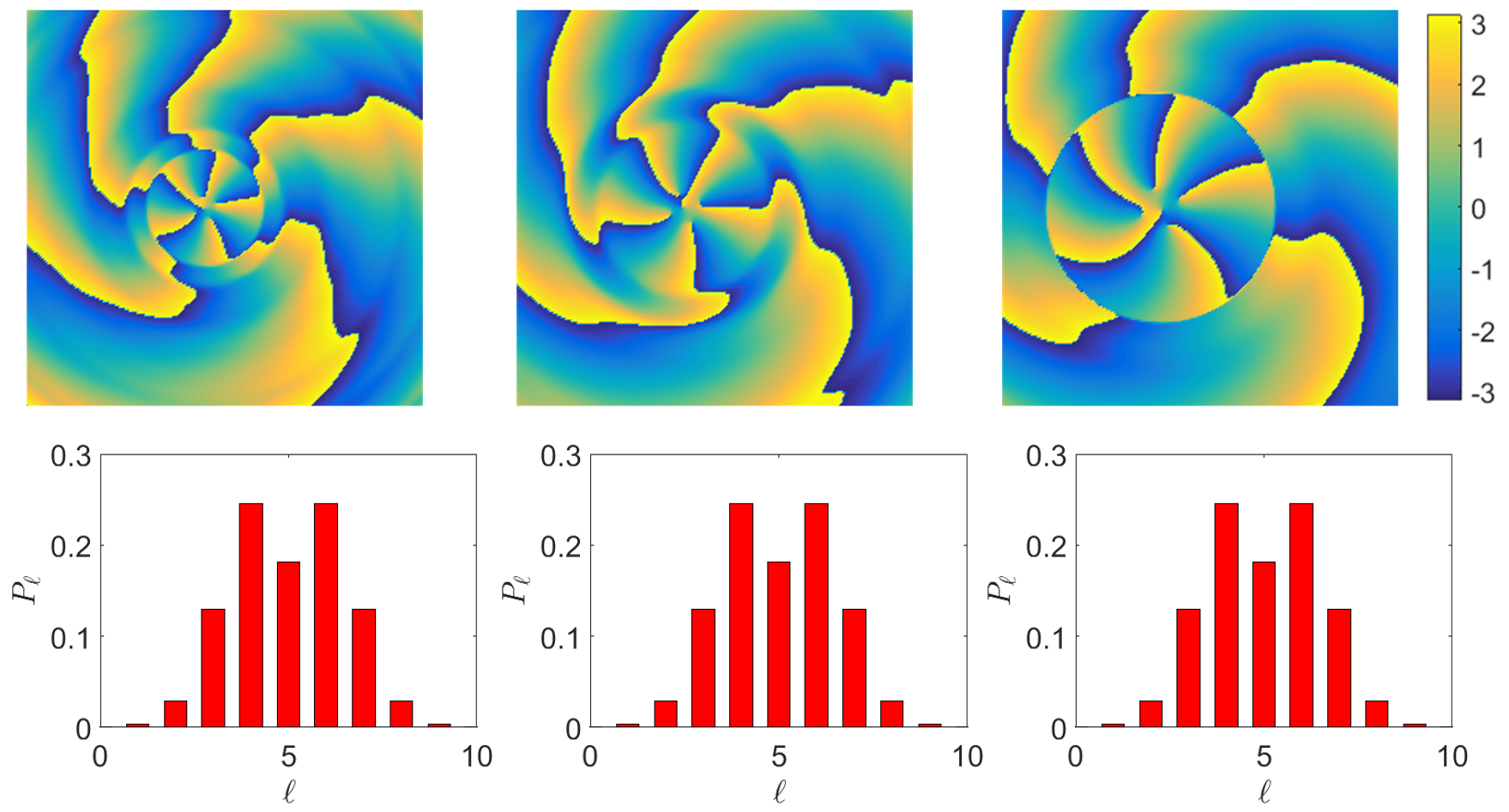}
	\caption{Phase profiles and angular spectra at the screen. The dimensions of all phase profiles are 1 mm $\times$ 1 mm, and $\ell=5$ and $z_1=0.5$ m are used. From left to right: $z_2$ = 0.1 m, 0.15 m, and 0.2 m.}
	\label{f7}
\end{figure}

Based upon this method, phase profiles and angular spectra at the screen are given in Fig. \ref{f7}.
Notice that these spectra are calculated by a large enough space, instead of the screen with diameter of 1 mm.
One can find asymmetric 5-fold phase jumps as a consequence of sloping incident field, which means a spread in angular spectrum.
Only when incident field is perpendicular to spiral phase is a pure spectrum without spread obtained. 
In addition, the angular spectrum stays the same with increasing of propagation distance.

%


\begin{thebibliography}{23}%
	\makeatletter
	\providecommand \@ifxundefined [1]{%
		\@ifx{#1\undefined}
	}%
	\providecommand \@ifnum [1]{%
		\ifnum #1\expandafter \@firstoftwo
		\else \expandafter \@secondoftwo
		\fi
	}%
	\providecommand \@ifx [1]{%
		\ifx #1\expandafter \@firstoftwo
		\else \expandafter \@secondoftwo
		\fi
	}%
	\providecommand \natexlab [1]{#1}%
	\providecommand \enquote  [1]{``#1''}%
	\providecommand \bibnamefont  [1]{#1}%
	\providecommand \bibfnamefont [1]{#1}%
	\providecommand \citenamefont [1]{#1}%
	\providecommand \href@noop [0]{\@secondoftwo}%
	\providecommand \href [0]{\begingroup \@sanitize@url \@href}%
	\providecommand \@href[1]{\@@startlink{#1}\@@href}%
	\providecommand \@@href[1]{\endgroup#1\@@endlink}%
	\providecommand \@sanitize@url [0]{\catcode `\\12\catcode `\$12\catcode
		`\&12\catcode `\#12\catcode `\^12\catcode `\_12\catcode `\%12\relax}%
	\providecommand \@@startlink[1]{}%
	\providecommand \@@endlink[0]{}%
	\providecommand \url  [0]{\begingroup\@sanitize@url \@url }%
	\providecommand \@url [1]{\endgroup\@href {#1}{\urlprefix }}%
	\providecommand \urlprefix  [0]{URL }%
	\providecommand \Eprint [0]{\href }%
	\providecommand \doibase [0]{http://dx.doi.org/}%
	\providecommand \selectlanguage [0]{\@gobble}%
	\providecommand \bibinfo  [0]{\@secondoftwo}%
	\providecommand \bibfield  [0]{\@secondoftwo}%
	\providecommand \translation [1]{[#1]}%
	\providecommand \BibitemOpen [0]{}%
	\providecommand \bibitemStop [0]{}%
	\providecommand \bibitemNoStop [0]{.\EOS\space}%
	\providecommand \EOS [0]{\spacefactor3000\relax}%
	\providecommand \BibitemShut  [1]{\csname bibitem#1\endcsname}%
	\let\auto@bib@innerbib\@empty
	\bibitem [{\citenamefont {Gariepy}\ \emph {et~al.}(2015)\citenamefont
		{Gariepy}, \citenamefont {Krstaji\'c}, \citenamefont {Henderson},
		\citenamefont {Li}, \citenamefont {Thomson}, \citenamefont {Buller},
		\citenamefont {Heshmat}, \citenamefont {Raskar}, \citenamefont {Leach},\ and\
		\citenamefont {Faccio}}]{Gariepy2015Single}%
	\BibitemOpen
	\bibfield  {author} {\bibinfo {author} {\bibfnamefont {G.}~\bibnamefont
			{Gariepy}}, \bibinfo {author} {\bibfnamefont {N.}~\bibnamefont {Krstaji\'c}},
		\bibinfo {author} {\bibfnamefont {R.}~\bibnamefont {Henderson}}, \bibinfo
		{author} {\bibfnamefont {C.}~\bibnamefont {Li}}, \bibinfo {author}
		{\bibfnamefont {R.~R.}\ \bibnamefont {Thomson}}, \bibinfo {author}
		{\bibfnamefont {G.~S.}\ \bibnamefont {Buller}}, \bibinfo {author}
		{\bibfnamefont {B.}~\bibnamefont {Heshmat}}, \bibinfo {author} {\bibfnamefont
			{R.}~\bibnamefont {Raskar}}, \bibinfo {author} {\bibfnamefont
			{J.}~\bibnamefont {Leach}}, \ and\ \bibinfo {author} {\bibfnamefont
			{D.}~\bibnamefont {Faccio}},\ }\href
	{https://www.nature.com/articles/ncomms7021} {\bibfield  {journal} {\bibinfo
			{journal} {Nat. Commun.}\ }\textbf {\bibinfo {volume} {6}},\ \bibinfo {pages}
		{6021} (\bibinfo {year} {2015})}\BibitemShut {NoStop}%
	\bibitem [{\citenamefont {Shin}\ \emph {et~al.}(2016)\citenamefont {Shin},
		\citenamefont {Xu}, \citenamefont {Venkatraman}, \citenamefont {Lussana},
		\citenamefont {Villa}, \citenamefont {Zappa}, \citenamefont {Goyal},
		\citenamefont {Wong},\ and\ \citenamefont {Shapiro}}]{Shin2016Photon}%
	\BibitemOpen
	\bibfield  {author} {\bibinfo {author} {\bibfnamefont {D.}~\bibnamefont
			{Shin}}, \bibinfo {author} {\bibfnamefont {F.}~\bibnamefont {Xu}}, \bibinfo
		{author} {\bibfnamefont {D.}~\bibnamefont {Venkatraman}}, \bibinfo {author}
		{\bibfnamefont {R.}~\bibnamefont {Lussana}}, \bibinfo {author} {\bibfnamefont
			{F.}~\bibnamefont {Villa}}, \bibinfo {author} {\bibfnamefont
			{F.}~\bibnamefont {Zappa}}, \bibinfo {author} {\bibfnamefont {V.~K.}\
			\bibnamefont {Goyal}}, \bibinfo {author} {\bibfnamefont {F.~N.~C.}\
			\bibnamefont {Wong}}, \ and\ \bibinfo {author} {\bibfnamefont {J.~H.}\
			\bibnamefont {Shapiro}},\ }\href
	{https://www.nature.com/articles/ncomms12046} {\bibfield  {journal} {\bibinfo
			{journal} {Nat. Commun.}\ }\textbf {\bibinfo {volume} {7}},\ \bibinfo
		{pages} {12046} (\bibinfo {year} {2016})}\BibitemShut {NoStop}%
	\bibitem [{\citenamefont {Ma}\ \emph {et~al.}(2012)\citenamefont {Ma},
		\citenamefont {Herbst}, \citenamefont {Scheidl}, \citenamefont {Wang},
		\citenamefont {Kropatschek}, \citenamefont {Naylor}, \citenamefont
		{Wittmann}, \citenamefont {Mech}, \citenamefont {Kofler},\ and\ \citenamefont
		{Anisimova}}]{Ma2012Quantum}%
	\BibitemOpen
	\bibfield  {author} {\bibinfo {author} {\bibfnamefont {X.~S.}\ \bibnamefont
			{Ma}}, \bibinfo {author} {\bibfnamefont {T.}~\bibnamefont {Herbst}}, \bibinfo
		{author} {\bibfnamefont {T.}~\bibnamefont {Scheidl}}, \bibinfo {author}
		{\bibfnamefont {D.}~\bibnamefont {Wang}}, \bibinfo {author} {\bibfnamefont
			{S.}~\bibnamefont {Kropatschek}}, \bibinfo {author} {\bibfnamefont
			{W.}~\bibnamefont {Naylor}}, \bibinfo {author} {\bibfnamefont
			{B.}~\bibnamefont {Wittmann}}, \bibinfo {author} {\bibfnamefont
			{A.}~\bibnamefont {Mech}}, \bibinfo {author} {\bibfnamefont {J.}~\bibnamefont
			{Kofler}}, \ and\ \bibinfo {author} {\bibfnamefont {E.}~\bibnamefont
			{Anisimova}},\ }\href {https://www.nature.com/articles/nature11472}
	{\bibfield  {journal} {\bibinfo  {journal} {Nature}\ }\textbf {\bibinfo
			{volume} {489}},\ \bibinfo {pages} {269} (\bibinfo {year}
		{2012})}\BibitemShut {NoStop}%
	\bibitem [{\citenamefont {Krenn}\ \emph {et~al.}(2016)\citenamefont {Krenn},
		\citenamefont {Handsteiner}, \citenamefont {Fink}, \citenamefont {Fickler},
		\citenamefont {Ursin}, \citenamefont {Malik},\ and\ \citenamefont
		{Zeilinger}}]{Krenn13648}%
	\BibitemOpen
	\bibfield  {author} {\bibinfo {author} {\bibfnamefont {M.}~\bibnamefont
			{Krenn}}, \bibinfo {author} {\bibfnamefont {J.}~\bibnamefont {Handsteiner}},
		\bibinfo {author} {\bibfnamefont {M.}~\bibnamefont {Fink}}, \bibinfo {author}
		{\bibfnamefont {R.}~\bibnamefont {Fickler}}, \bibinfo {author} {\bibfnamefont
			{R.}~\bibnamefont {Ursin}}, \bibinfo {author} {\bibfnamefont
			{M.}~\bibnamefont {Malik}}, \ and\ \bibinfo {author} {\bibfnamefont
			{A.}~\bibnamefont {Zeilinger}},\ }\href {\doibase 10.1073/pnas.1612023113}
	{\bibfield  {journal} {\bibinfo  {journal} {Proc. Natl. Acad. Sci.}\ }\textbf
		{\bibinfo {volume} {113}},\ \bibinfo {pages} {13648} (\bibinfo {year}
		{2016})}\BibitemShut {NoStop}%
	\bibitem [{\citenamefont {van~den Berg}\ \emph {et~al.}(2012)\citenamefont
		{van~den Berg}, \citenamefont {Persijn}, \citenamefont {Kok}, \citenamefont
		{Zeitouny},\ and\ \citenamefont {Bhattacharya}}]{PhysRevLett.108.183901}%
	\BibitemOpen
	\bibfield  {author} {\bibinfo {author} {\bibfnamefont {S.~A.}\ \bibnamefont
			{van~den Berg}}, \bibinfo {author} {\bibfnamefont {S.~T.}\ \bibnamefont
			{Persijn}}, \bibinfo {author} {\bibfnamefont {G.~J.~P.}\ \bibnamefont {Kok}},
		\bibinfo {author} {\bibfnamefont {M.~G.}\ \bibnamefont {Zeitouny}}, \ and\
		\bibinfo {author} {\bibfnamefont {N.}~\bibnamefont {Bhattacharya}},\ }\href
	{\doibase 10.1103/PhysRevLett.108.183901} {\bibfield  {journal} {\bibinfo
			{journal} {Phys. Rev. Lett.}\ }\textbf {\bibinfo {volume} {108}},\ \bibinfo
		{pages} {183901} (\bibinfo {year} {2012})}\BibitemShut {NoStop}%
	\bibitem [{\citenamefont {Lavery}\ \emph {et~al.}(2013)\citenamefont {Lavery},
		\citenamefont {Speirits}, \citenamefont {Barnett},\ and\ \citenamefont
		{Padgett}}]{Lavery537}%
	\BibitemOpen
	\bibfield  {author} {\bibinfo {author} {\bibfnamefont {M.~P.~J.}\
			\bibnamefont {Lavery}}, \bibinfo {author} {\bibfnamefont {F.~C.}\
			\bibnamefont {Speirits}}, \bibinfo {author} {\bibfnamefont {S.~M.}\
			\bibnamefont {Barnett}}, \ and\ \bibinfo {author} {\bibfnamefont {M.~J.}\
			\bibnamefont {Padgett}},\ }\href {\doibase 10.1126/science.1239936}
	{\bibfield  {journal} {\bibinfo  {journal} {Science}\ }\textbf {\bibinfo
			{volume} {341}},\ \bibinfo {pages} {537} (\bibinfo {year}
		{2013})}\BibitemShut {NoStop}%
	\bibitem [{\citenamefont {Pan}\ \emph {et~al.}(2012)\citenamefont {Pan},
		\citenamefont {Chen}, \citenamefont {Lu}, \citenamefont {Weinfurter},
		\citenamefont {Zeilinger},\ and\ \citenamefont {\ifmmode~\dot{Z}\else
			\.{Z}\fi{}ukowski}}]{RevModPhys.84.777}%
	\BibitemOpen
	\bibfield  {author} {\bibinfo {author} {\bibfnamefont {J.-W.}\ \bibnamefont
			{Pan}}, \bibinfo {author} {\bibfnamefont {Z.-B.}\ \bibnamefont {Chen}},
		\bibinfo {author} {\bibfnamefont {C.-Y.}\ \bibnamefont {Lu}}, \bibinfo
		{author} {\bibfnamefont {H.}~\bibnamefont {Weinfurter}}, \bibinfo {author}
		{\bibfnamefont {A.}~\bibnamefont {Zeilinger}}, \ and\ \bibinfo {author}
		{\bibfnamefont {M.}~\bibnamefont {\ifmmode~\dot{Z}\else \.{Z}\fi{}ukowski}},\
	}\href {\doibase 10.1103/RevModPhys.84.777} {\bibfield  {journal} {\bibinfo
			{journal} {Rev. Mod. Phys.}\ }\textbf {\bibinfo {volume} {84}},\ \bibinfo
		{pages} {777} (\bibinfo {year} {2012})}\BibitemShut {NoStop}%
	\bibitem [{\citenamefont {Lloyd}(2008)}]{Lloyd1463}%
	\BibitemOpen
	\bibfield  {author} {\bibinfo {author} {\bibfnamefont {S.}~\bibnamefont
			{Lloyd}},\ }\href {\doibase 10.1126/science.1160627} {\bibfield  {journal}
		{\bibinfo  {journal} {Science}\ }\textbf {\bibinfo {volume} {321}},\ \bibinfo
		{pages} {1463} (\bibinfo {year} {2008})}\BibitemShut {NoStop}%
	\bibitem [{\citenamefont {England}\ \emph {et~al.}(2019)\citenamefont
		{England}, \citenamefont {Balaji},\ and\ \citenamefont
		{Sussman}}]{PhysRevA.99.023828}%
	\BibitemOpen
	\bibfield  {author} {\bibinfo {author} {\bibfnamefont {D.~G.}\ \bibnamefont
			{England}}, \bibinfo {author} {\bibfnamefont {B.}~\bibnamefont {Balaji}}, \
		and\ \bibinfo {author} {\bibfnamefont {B.~J.}\ \bibnamefont {Sussman}},\
	}\href {\doibase 10.1103/PhysRevA.99.023828} {\bibfield  {journal} {\bibinfo
			{journal} {Phys. Rev. A}\ }\textbf {\bibinfo {volume} {99}},\ \bibinfo
		{pages} {023828} (\bibinfo {year} {2019})}\BibitemShut {NoStop}%
	\bibitem [{\citenamefont {Lopaeva}\ \emph {et~al.}(2013)\citenamefont
		{Lopaeva}, \citenamefont {Ruo~Berchera}, \citenamefont {Degiovanni},
		\citenamefont {Olivares}, \citenamefont {Brida},\ and\ \citenamefont
		{Genovese}}]{PhysRevLett.110.153603}%
	\BibitemOpen
	\bibfield  {author} {\bibinfo {author} {\bibfnamefont {E.~D.}\ \bibnamefont
			{Lopaeva}}, \bibinfo {author} {\bibfnamefont {I.}~\bibnamefont
			{Ruo~Berchera}}, \bibinfo {author} {\bibfnamefont {I.~P.}\ \bibnamefont
			{Degiovanni}}, \bibinfo {author} {\bibfnamefont {S.}~\bibnamefont
			{Olivares}}, \bibinfo {author} {\bibfnamefont {G.}~\bibnamefont {Brida}}, \
		and\ \bibinfo {author} {\bibfnamefont {M.}~\bibnamefont {Genovese}},\ }\href
	{\doibase 10.1103/PhysRevLett.110.153603} {\bibfield  {journal} {\bibinfo
			{journal} {Phys. Rev. Lett.}\ }\textbf {\bibinfo {volume} {110}},\ \bibinfo
		{pages} {153603} (\bibinfo {year} {2013})}\BibitemShut {NoStop}%
	\bibitem [{\citenamefont {Gregory}\ \emph {et~al.}(2019)\citenamefont
		{Gregory}, \citenamefont {Moreau}, \citenamefont {Toninelli},\ and\
		\citenamefont {Padgett}}]{gregory2019imaging}%
	\BibitemOpen
	\bibfield  {author} {\bibinfo {author} {\bibfnamefont {T.}~\bibnamefont
			{Gregory}}, \bibinfo {author} {\bibfnamefont {P.-A.}\ \bibnamefont {Moreau}},
		\bibinfo {author} {\bibfnamefont {E.}~\bibnamefont {Toninelli}}, \ and\
		\bibinfo {author} {\bibfnamefont {M.~J.}\ \bibnamefont {Padgett}},\
	}\href@noop {} {\bibfield  {journal} {\bibinfo  {journal} {arXiv preprint
				arXiv:1907.09370}\ } (\bibinfo {year} {2019})}\BibitemShut {NoStop}%
	\bibitem [{\citenamefont {Malik}\ \emph {et~al.}(2012)\citenamefont {Malik},
		\citenamefont {Maga\~na Loaiza},\ and\ \citenamefont
		{Boyd}}]{doi:10.1063/1.4770298}%
	\BibitemOpen
	\bibfield  {author} {\bibinfo {author} {\bibfnamefont {M.}~\bibnamefont
			{Malik}}, \bibinfo {author} {\bibfnamefont {O.~S.}\ \bibnamefont {Maga\~na
				Loaiza}}, \ and\ \bibinfo {author} {\bibfnamefont {R.~W.}\ \bibnamefont
			{Boyd}},\ }\href {\doibase 10.1063/1.4770298} {\bibfield  {journal} {\bibinfo
			{journal} {Appl. Phys. Lett.}\ }\textbf {\bibinfo {volume} {101}},\ \bibinfo
		{pages} {241103} (\bibinfo {year} {2012})}\BibitemShut {NoStop}%
	\bibitem [{\citenamefont {Morris}\ \emph {et~al.}(2015)\citenamefont {Morris},
		\citenamefont {Aspden}, \citenamefont {Bell}, \citenamefont {Boyd},\ and\
		\citenamefont {Padgett}}]{Morris2015Imaging}%
	\BibitemOpen
	\bibfield  {author} {\bibinfo {author} {\bibfnamefont {P.~A.}\ \bibnamefont
			{Morris}}, \bibinfo {author} {\bibfnamefont {R.~S.}\ \bibnamefont {Aspden}},
		\bibinfo {author} {\bibfnamefont {J.~E.}\ \bibnamefont {Bell}}, \bibinfo
		{author} {\bibfnamefont {R.~W.}\ \bibnamefont {Boyd}}, \ and\ \bibinfo
		{author} {\bibfnamefont {M.~J.}\ \bibnamefont {Padgett}},\ }\href
	{https://www.nature.com/articles/ncomms6913} {\bibfield  {journal} {\bibinfo
			{journal} {Nat. Commun.}\ }\textbf {\bibinfo {volume} {6}},\ \bibinfo {pages}
		{5913} (\bibinfo {year} {2015})}\BibitemShut {NoStop}%
	\bibitem [{\citenamefont {Li}\ \emph {et~al.}(2019)\citenamefont {Li},
		\citenamefont {Huang}, \citenamefont {Cao}, \citenamefont {Wang},
		\citenamefont {Li}, \citenamefont {Jin}, \citenamefont {Yu}, \citenamefont
		{Zhang}, \citenamefont {Zhang}, \citenamefont {Peng}, \citenamefont {Xu},\
		and\ \citenamefont {Pan}}]{li2019single}%
	\BibitemOpen
	\bibfield  {author} {\bibinfo {author} {\bibfnamefont {Z.-P.}\ \bibnamefont
			{Li}}, \bibinfo {author} {\bibfnamefont {X.}~\bibnamefont {Huang}}, \bibinfo
		{author} {\bibfnamefont {Y.}~\bibnamefont {Cao}}, \bibinfo {author}
		{\bibfnamefont {B.}~\bibnamefont {Wang}}, \bibinfo {author} {\bibfnamefont
			{Y.-H.}\ \bibnamefont {Li}}, \bibinfo {author} {\bibfnamefont
			{W.}~\bibnamefont {Jin}}, \bibinfo {author} {\bibfnamefont {C.}~\bibnamefont
			{Yu}}, \bibinfo {author} {\bibfnamefont {J.}~\bibnamefont {Zhang}}, \bibinfo
		{author} {\bibfnamefont {Q.}~\bibnamefont {Zhang}}, \bibinfo {author}
		{\bibfnamefont {C.-Z.}\ \bibnamefont {Peng}}, \bibinfo {author}
		{\bibfnamefont {F.}~\bibnamefont {Xu}}, \ and\ \bibinfo {author}
		{\bibfnamefont {J.-W.}\ \bibnamefont {Pan}},\ }\href@noop {} {\bibfield
		{journal} {\bibinfo  {journal} {arXiv preprint arXiv:1904.10341}\ } (\bibinfo
		{year} {2019})}\BibitemShut {NoStop}%
	\bibitem [{\citenamefont {Cohen}\ \emph {et~al.}(2019)\citenamefont {Cohen},
		\citenamefont {Matekole}, \citenamefont {Sher}, \citenamefont {Istrati},
		\citenamefont {Eisenberg},\ and\ \citenamefont
		{Dowling}}]{cohen2019thresholded}%
	\BibitemOpen
	\bibfield  {author} {\bibinfo {author} {\bibfnamefont {L.}~\bibnamefont
			{Cohen}}, \bibinfo {author} {\bibfnamefont {E.~S.}\ \bibnamefont {Matekole}},
		\bibinfo {author} {\bibfnamefont {Y.}~\bibnamefont {Sher}}, \bibinfo {author}
		{\bibfnamefont {D.}~\bibnamefont {Istrati}}, \bibinfo {author} {\bibfnamefont
			{H.~S.}\ \bibnamefont {Eisenberg}}, \ and\ \bibinfo {author} {\bibfnamefont
			{J.~P.}\ \bibnamefont {Dowling}},\ }\href@noop {} {\bibfield  {journal}
		{\bibinfo  {journal} {arXiv preprint arXiv:1906.09615}\ } (\bibinfo {year}
		{2019})}\BibitemShut {NoStop}%
	\bibitem [{\citenamefont {Dennis}\ \emph {et~al.}(2009)\citenamefont {Dennis},
		\citenamefont {O'Holleran},\ and\ \citenamefont {Padgett}}]{DENNIS2009293}%
	\BibitemOpen
	\bibfield  {author} {\bibinfo {author} {\bibfnamefont {M.~R.}\ \bibnamefont
			{Dennis}}, \bibinfo {author} {\bibfnamefont {K.}~\bibnamefont {O'Holleran}},
		\ and\ \bibinfo {author} {\bibfnamefont {M.~J.}\ \bibnamefont {Padgett}}\
	}(\bibinfo  {publisher} {Elsevier},\ \bibinfo {year} {2009})\ pp.\ \bibinfo
	{pages} {293 -- 363}\BibitemShut {NoStop}%
	\bibitem [{not()}]{note1}%
	\BibitemOpen
	\href@noop {} {}\bibinfo {note} {The signal amplitude after SLM1 is a
		superposition of LG modes, ${E_{\rm S}} = \sum\nolimits_p \alpha_p
		{{E_{\ell,p}}}$, with the same azimuthal index. The LG mode with $p=0$
		appears in the inner when compared to those with $p \ne 0$; meanwhile, there
		is a small ingredient of LG modes with non-zero $p$, as shown in Ref.
		\cite{DENNIS2009293}. Therfore, the signal intensity can be written as
		${I_{\rm S}} = \eta {I_{\ell,p = 0}} + \left( {1 - \eta } \right){I_{\ell,p
				\ne 0}}$ with a large weight $\eta = \left| \alpha_0 \right| ^2$. Herein we
		merely take the first term since signle-mode fiber can filter these
		modes.}\BibitemShut {Stop}%
	\bibitem [{\citenamefont {Allen}\ \emph {et~al.}(1992)\citenamefont {Allen},
		\citenamefont {Beijersbergen}, \citenamefont {Spreeuw},\ and\ \citenamefont
		{Woerdman}}]{PhysRevA.45.8185}%
	\BibitemOpen
	\bibfield  {author} {\bibinfo {author} {\bibfnamefont {L.}~\bibnamefont
			{Allen}}, \bibinfo {author} {\bibfnamefont {M.~W.}\ \bibnamefont
			{Beijersbergen}}, \bibinfo {author} {\bibfnamefont {R.~J.~C.}\ \bibnamefont
			{Spreeuw}}, \ and\ \bibinfo {author} {\bibfnamefont {J.~P.}\ \bibnamefont
			{Woerdman}},\ }\href {\doibase 10.1103/PhysRevA.45.8185} {\bibfield
		{journal} {\bibinfo  {journal} {Phys. Rev. A}\ }\textbf {\bibinfo {volume}
			{45}},\ \bibinfo {pages} {8185} (\bibinfo {year} {1992})}\BibitemShut
	{NoStop}%
	\bibitem [{\citenamefont {Palacios}\ \emph {et~al.}(2004)\citenamefont
		{Palacios}, \citenamefont {Maleev}, \citenamefont {Marathay},\ and\
		\citenamefont {Swartzlander}}]{PhysRevLett.92.143905}%
	\BibitemOpen
	\bibfield  {author} {\bibinfo {author} {\bibfnamefont {D.~M.}\ \bibnamefont
			{Palacios}}, \bibinfo {author} {\bibfnamefont {I.~D.}\ \bibnamefont
			{Maleev}}, \bibinfo {author} {\bibfnamefont {A.~S.}\ \bibnamefont
			{Marathay}}, \ and\ \bibinfo {author} {\bibfnamefont {G.~A.}\ \bibnamefont
			{Swartzlander}},\ }\href {\doibase 10.1103/PhysRevLett.92.143905} {\bibfield
		{journal} {\bibinfo  {journal} {Phys. Rev. Lett.}\ }\textbf {\bibinfo
			{volume} {92}},\ \bibinfo {pages} {143905} (\bibinfo {year}
		{2004})}\BibitemShut {NoStop}%
	\bibitem [{\citenamefont {Hetharia}\ \emph {et~al.}(2014)\citenamefont
		{Hetharia}, \citenamefont {van Exter},\ and\ \citenamefont
		{L\"offler}}]{PhysRevA.90.063801}%
	\BibitemOpen
	\bibfield  {author} {\bibinfo {author} {\bibfnamefont {D.}~\bibnamefont
			{Hetharia}}, \bibinfo {author} {\bibfnamefont {M.~P.}\ \bibnamefont {van
				Exter}}, \ and\ \bibinfo {author} {\bibfnamefont {W.}~\bibnamefont
			{L\"offler}},\ }\href {\doibase 10.1103/PhysRevA.90.063801} {\bibfield
		{journal} {\bibinfo  {journal} {Phys. Rev. A}\ }\textbf {\bibinfo {volume}
			{90}},\ \bibinfo {pages} {063801} (\bibinfo {year} {2014})}\BibitemShut
	{NoStop}%
	\bibitem [{\citenamefont {Swartzlander}\ and\ \citenamefont
		{Hernandez-Aranda}(2007)}]{PhysRevLett.99.163901}%
	\BibitemOpen
	\bibfield  {author} {\bibinfo {author} {\bibfnamefont {G.~A.}\ \bibnamefont
			{Swartzlander}}\ and\ \bibinfo {author} {\bibfnamefont {R.~I.}\ \bibnamefont
			{Hernandez-Aranda}},\ }\href {\doibase 10.1103/PhysRevLett.99.163901}
	{\bibfield  {journal} {\bibinfo  {journal} {Phys. Rev. Lett.}\ }\textbf
		{\bibinfo {volume} {99}},\ \bibinfo {pages} {163901} (\bibinfo {year}
		{2007})}\BibitemShut {NoStop}%
	\bibitem [{\citenamefont {Zhang}\ \emph {et~al.}(2019)\citenamefont {Zhang},
		\citenamefont {Zhang}, \citenamefont {Hu}, \citenamefont {Cen}, \citenamefont
		{Sun}, \citenamefont {Jin},\ and\ \citenamefont {Zhao}}]{zhang2019angular}%
	\BibitemOpen
	\bibfield  {author} {\bibinfo {author} {\bibfnamefont {J.-D.}\ \bibnamefont
			{Zhang}}, \bibinfo {author} {\bibfnamefont {Z.-J.}\ \bibnamefont {Zhang}},
		\bibinfo {author} {\bibfnamefont {J.-Y.}\ \bibnamefont {Hu}}, \bibinfo
		{author} {\bibfnamefont {L.-Z.}\ \bibnamefont {Cen}}, \bibinfo {author}
		{\bibfnamefont {Y.-F.}\ \bibnamefont {Sun}}, \bibinfo {author} {\bibfnamefont
			{C.-F.}\ \bibnamefont {Jin}}, \ and\ \bibinfo {author} {\bibfnamefont
			{Y.}~\bibnamefont {Zhao}},\ }\href@noop {} {\bibfield  {journal} {\bibinfo
			{journal} {arXiv preprint arXiv:1906.08474}\ } (\bibinfo {year}
		{2019})}\BibitemShut {NoStop}%
	\bibitem [{\citenamefont {Paterson}(2005)}]{PhysRevLett.94.153901}%
	\BibitemOpen
	\bibfield  {author} {\bibinfo {author} {\bibfnamefont {C.}~\bibnamefont
			{Paterson}},\ }\href {\doibase 10.1103/PhysRevLett.94.153901} {\bibfield
		{journal} {\bibinfo  {journal} {Phys. Rev. Lett.}\ }\textbf {\bibinfo
			{volume} {94}},\ \bibinfo {pages} {153901} (\bibinfo {year}
		{2005})}\BibitemShut {NoStop}%
\end{thebibliography}
\end{document}